\documentclass{elsarticle}

\usepackage{lineno,hyperref}
\modulolinenumbers[5]

\usepackage{etex}
\usepackage{times}

 \usepackage[scaled=.9]{helvet}

\usepackage{latexsym}

\usepackage{listings}
\lstdefinelanguage{pseudocode}{
	morekeywords = {repeat, until, if, else, wait, send, receive, unless, while, for, return, break},
	sensitive = false,
	morecomment=[l]{//},
	morecomment=[s]{/*}{*/},
}
\usepackage[noend]{algpseudocode}
\usepackage{etex}
\usepackage{mathptmx}
\usepackage{graphicx}
\usepackage{latexsym}
\usepackage{amsmath,amssymb}

\newcommand{\Y}{\mathbb{Y}}

\reserveinserts{28}

\newtheorem{theorem}{Theorem}
\newtheorem{lemma}[theorem]{Lemma}

\newtheorem{corollary}{Corollary}

\newcommand{\sens}{\mathit{sens}}

\newcommand{\ignore}[1] {}

\newcommand{\abs}[1]{| #1 |}

\begin{document}

\begin{frontmatter}

\title{On testing substitutability }

\author[Caddress]{Cosmina Croitoru}
\author[Kaddress]{Kurt Mehlhorn}

\address[Caddress]{Saarland University, Saarbr\"ucken}
\address[Kaddress]{Max Planck Institut for Informatics, Saarbr\"ucken}


\begin{abstract}
The papers~\cite{hatfimmokomi11} and~\cite{azizbrilharr13} propose
algorithms for testing whether the choice function induced by  a (strict) preference list of length $N$ over a universe $U$ is substitutable. The running time of these algorithms is $O(|U|^3\cdot N^3)$, respectively  $O(|U|^2\cdot N^3)$. In this note we present an algorithm with running time 
$O(|U|^2\cdot N^2)$. Note that $N$ may be exponential in the size $|U|$ of the universe.

\end{abstract}

\end{frontmatter}

A \emph{choice function} on a finite set $U$ of alternatives is any function $f$ from subsets of $U$ to subsets of $U$ that maps any set $A$ to a subset of itself, i.e., $f(A) \subseteq A$ for all $A \subseteq U$. A choice function $f$ is \emph{substitutable} if 
\[    \text{$A \subseteq B$ implies $f(B) \cap A \subseteq f(A)$ for all $A,B \subseteq U$,}\]
i.e. the additional alternatives provided by $B$ do not promote any $x \in A - f(A)$ to the set of selected elements. 

We are interested in choice functions induced by preference lists $\Y$ on subsets of $U$. A preference list $\Y$ is simply an ordered list of subsets of $U$ and the associated choice function $f_{\Y}$ maps any subset $A$ of $U$ to the first element on the list that is contained in $A$. If $\Y$ is understood from the context, we write $f$ instead of $f_{\Y}$.
 We use $N$ to denote the number of elements on $\Y$, and, in order to make $f$ defined for all $A$, we assume that the empty set is the last element of $\Y$.
For elements $X$ and $Y$ in $\Y$, we write $X \succ Y$ if $X$ properly precedes $Y$ on $\Y$ and we write $X \succeq Y$ for $X \succ Y$ or $X = Y$. 

For example, let $U = \{a,b,c,d\}$ and $\Y\hspace{-1pt} = (\{a,b\},\{a,c,d\}, \{a,c\}, \{a\},\{c\},\emptyset)$. Then $f_{\Y}(\{a,b,c\}) = \{a,b\}$.  The function 
$f_{\Y}$ is not substitutable since $d \in (f_{\Y}(\{a,c,d\}) \cap \{d\}) - f_{\Y}(\{d\})$. 
We refer to~\cite{hatfimmokomi11} for a discussion of the role of substitutable choice functions in economics. 

$\Y$ is \emph{coherent} if $X \succ Y$ implies $X \not\subseteq Y$ for any two elements on $\Y$. Assume $X \succ Y$ and $X \subseteq Y$. Then $Y$ does not lie in the range of $f_{\Y}$ and removing $Y$ from $\Y$ does not change the function $f$. Thus we may assume that $\Y$ is coherent. 

From now on, $\Y$ denotes a coherent preference list and $f$ stands for $f_{\Y}$. $\Y$ is \textit{substitutable} if 
$f$ is a substitutable choice function.

\begin{lemma} Let $\Y$ be a coherent preference list on $U$. Then for any $A \subseteq U$, $f(A) = A$ if and only if $A \in \Y$. \end{lemma}
\textit{Proof.} Since $f$ maps the powerset of $U$ to $\Y$, $f(A) = A$ implies $A \in \Y$. Conversely, assume $A \in \Y$ and $f(A) \succ A$. Then $f(A)$ and $A$ are members of $\Y$ with $f(A) \succ A$ and $f(A) \subseteq A$, a contradiction to the coherence of $\Y$. \hfill $\Box$ 

 An established condition of choice functions known as Aizerman's  \textit{outcast},
 or  Chernoff's \textit{postulate $5^*$}\,, or $\hat{\alpha}$ 
  (see Brandt and  Harrenstein  \cite{branharre11})
  is 
 \[   (outcast):\quad\quad \text{if $f(A) \subseteq B \subseteq A$ then $f(B) = f(A)$.}\]

\begin{lemma} If $Y$ is a coherent preference list on $U$, then $f$ satisfies \emph{outcast}. 
˚
\end{lemma}
\textit{Proof.} $B \subseteq A$ implies $f(A) \succeq f(B)$ and $f(A) \subseteq B$ implies $f(B) \succeq f(f(A)) = f(A)$, where the last equality uses coherence. Thus $f(A) = f(B)$.  \hfill $\Box$ 

\begin{lemma} Let $\Y$ be a coherent and substitutable preference list on $U$. If $X$ is a member of $\Y$ then also every subset of $X$ is a member of\,  $\Y$.
 \end{lemma}
\textit{Proof.} Assume $X = f(X)$ and $A \subseteq X$. By substitutability, $f(X) \cap A \subseteq f(A)$ and hence 
$A = X \cap A = f(X) \cap A \subseteq f(A)$. Thus $f(A) = A$.  \hfill $\Box$ 

\vskip0.2cm
A preference list $\Y$ is \textit{complete} if it contains for each $X \in Y$ also all of its subsets. Note that complete preference lists  are exponentially long in the size of their largest member. 

\vskip0.1cm

In order to demonstrate non-substitutability of a preference list, we need to exhibit sets $A$ and $B$ with 
$A \subseteq B$ and $f(B) \cap (A - f(A)) \not= \emptyset$. We next show that we can restrict the search to special subsets of $U$. A \emph{witness (to non-substitutability)} is a pair $(X,Y)$ of members of $\Y$ such that $X \succ Y$, $f(X \cup Y) = X$ and there is an $x \in X - Y$ such that $f(Y \cup \{x\}) = Y$. Note that $x$ is selected when the set of alternatives is $X \cup Y$ (this is the set $B$) but is not selected when the set of alternatives is $Y \cup \{x\}$ (this is the set $A$). 

\begin{theorem}\label{prop:thm1} \hspace{-4pt}$\Y$ is not substitutable if and only if there is a witness to non-substitutability. \end{theorem}
\textit{Proof.} Assume first that $(X,Y)$ is a witness. Then $X \succ Y$, $f(X \cup Y) = X$ and there is an $x \in X - Y$ such that $f(Y \cup x) = Y$. Let $A = Y \cup \{x\}$ and $B = X \cup Y$. Then $A \subseteq B$ and $x \in f(B) \cap (A - f(A))$. Thus $f$ is not substitutable. 

Conversely, assume that $f$ is not substitutable. Then there are subsets $A$ and $B$ of $U$ with $A \subseteq B$ and $f(B) \cap A \not \subseteq f(A)$. Since $A \subseteq B$, we have $f(B) \succeq f(A)$. In fact, $f(B) \succ f(A)$ since $f(B) = f(A)$ and $f(A) \subseteq A$ implies $f(B) \cap A = f(A)$. Since $f(A) \subseteq A \subseteq B$, we have
$f(A) \cup f(B) \subseteq B$ and hence $f(B) \subseteq f(A) \cup f(B) \subseteq B$. Thus $f(f(A) \cup f(B)) = f(B)$ by property (outcast). Let $x \in  (f(B) \cap A) - f(A)$. Then $f(B) \cup \{x\} \subseteq A$ and $f(A) \subseteq f(A) \cup \{x \}\subseteq A$ and hence $f(f(A) \cup \{x\}) = f(A)$ by (outcast). Thus $(f(B), f(A))$ is a witness.  \hfill $\Box$

Theorem~\ref{prop:thm1} directly translates into an algorithm of running time $O(N^3\abs{U} + N^2 \abs{U}^2)$. Note first that one can determine $f(A)$ in time $O(N \abs{U})$ by simply scanning the list $\Y$ and checking each set for containment. The algorithm has two phases. In the first phase, one determines for each $Y \in \Y$ the set of $x$ for which $f(Y \cup \{x\}) = Y$. This requires $N \abs{U}$ function evaluations and $O(N^2 \abs{U}^2)$ time. Then one checks for every pair $(X,Y)$ of elements of $\Y$, whether it is a witness. This requires $N^2$ function evaluations and $N^2 \abs{U}$ look-ups of precomputed values and hence takes time $O(N^3 \abs{U})$. 

We improve the running time to $O(N^2 \abs{U}^2)$. The crucial insight is as follows. We search for a witness pair $(X,Y)$ in increasing order of $X$. Of course, we stop the search as soon as we have found a witness. So when we consider a pair $(X,Y)$ we know that there is no witness $(Z,\cdot )$ with $Z \succ X$. We then have $f(X \cup Y) = X$ if and only if $f(X \cup \{x\}) = X$ for all elements $x \in Y -  X$. We stress that this equivalence does not hold in general, it only holds under the assumption that there is no earlier witness. So we can replace the function evaluation $f(X \cup Y)$ of cost $O(N \abs{U})$ by $\abs{U}$ look-ups of precomputed values. We next give the details.

We call $X \in \Y$ \emph{insensitive} to $x \in U$ if $f(X \cup \{x\}) = X$ and \emph{sensitive} otherwise. 

\begin{lemma}\label{key lemma} Let $X, Y \in \Y$ with $X \succ Y$. If $f(X \cup Y) = X$, then $X$ is insensitive to all $x \in Y - X$. If $X$ is insensitive to all $x \in Y - X$ and there is no witness $(Z,\cdot )$ with $Z \succ X$, then $f(X \cup Y) = X$. \end{lemma}
\textit{Proof.} Let $x \in Y - X$ be arbitrary. Then $X \subseteq X \cup \{x\} \subseteq X \cup Y$ and hence 
$X = f(X \cup Y) \succeq f(X \cup \{x\}) \succeq f(X) = X$. Thus $f(X \cup \{x\}) = X$ and $X$ is insensitive to $x$. 

For the second part, assume $f(X \cup Y) = Z$ with $Z \succ X$. Then $Z \subseteq X \cup Y$ and hence $Z \cup X \subseteq X \cup Y$. Thus $Z \succeq f(X \cup Y) \succeq f(X \cup Z) \succeq Z$, where the last inequality follows from $Z \subseteq X \cup Z$. Thus $f(X \cup Z) = Z$. Since $(Z,X)$ is not a witness, we must have
$f(X \cup \{x\}) \not= X$ for every $x \in Z - X$. On the other hand, $Z - X \subseteq Y - X$ (since $Z \subseteq X \cup Y$) and $f(X \cup \{x\}) = X$ since $X$ is insensitive to all $x \in Y - X$, a contradiction.  \hfill $\Box$

\vskip0.2cm

Lemma~\ref{key lemma} suggests a way to find the non-substitutability witness $(X,\cdot )$ with minimal first component. 

\newcommand{\false}{\mathit{false}}
\newcommand{\true}{\mathit{true}}

\begin{figure}[t]
\begin{lstlisting}
--------------------------------------------------------------------
$\mathbf{1.} \textbf{ Preprocessing}$
   $\textbf{for all }$ $X\in \Y$  $\textbf{do}$ $\{\ d_X:=1;$ $\textbf{for all }$ $x\in U$ $\textbf{do}$ $\sens(x,X) := \false \ \}$
   $\textbf{for all }$ $X\in \Y$ $\textbf{do}$
        $\textbf{for all }$ $Y\in \Y$ $\text{with }$ $X \succ Y$ $\textbf{do}\ \{$
            $\textbf{if}$ $X \subseteq Y$ $\textbf{then \ return } \mathbb{Y} \text{ is NOT COHERENT}$;
            $\textbf{if}$ $Y \subseteq X$ $\textbf{then }$ $\text{increment }  d_X\ $; 
            $\textbf{for all }$ $x\in U-Y \textbf{ do }$ $\textbf{if }$ $X \subseteq Y \cup \{x\}$ $ \textbf{then } \sens(x,Y) := \true$ $\}$            
   $\textbf{for all }$ $X\in \Y$  $\textbf{do}$  $\textbf{if}$ $d_X\neq 2^{|X|}$ $\textbf{then return } \mathbb{Y} \text{ is NOT COMPLETE}$;
   
$\mathbf{2.} \textbf{ Looking for the first witness to non-substitutability }$
   $\textbf{for all }$ $X\in \Y$ $\textbf{do}$
      $\textbf{for all }$ $Y\in \Y$ $\text{with }$ $X \succ Y$ $\textbf{do}\ $
          $\textbf{if}$ $(\exists x\in X-Y\, \text{s.t. } \sens(x,Y) = \true)\land (\forall y\in Y-X\   \sens(y,X)=\false)$ $\textbf{then }$
            $\textbf{return } \mathbb{Y} \text{ is NOT SUBSTITUTABLE: \text{ witness } } (X,Y) $;
            
$\mathbf{3.} \textbf{ Success}$ 
   $\textbf{return } \mathbb{Y} \text{ is  SUBSTITUTABLE}$          
---------------------------------------------------------------------          
\end{lstlisting} 

\vskip-0.3cm
 \caption{ $ \textbf{Testing if the list } \mathbb{Y} \textbf{ is substitutable}$
}\label{fig:as1}
\end{figure}

\begin{theorem}\label{main theorem}  Let $X, Y \in \Y$ with $X \succ Y$ and assume that there is no  witness $(Z,\cdot )$ with $Z \succ X$. Then $(X,Y)$ is a witness if and only if $X$ is insensitive to all $x \in Y - X$ and $Y$ is sensitive to some $x \in X - Y$. 
\end{theorem}
\textit{Proof.} Assume first that $(X,Y)$ is a witness pair. Then $Y$ is sensitive to some $x \in X - Y$ and $f(X \cup Y) = X$. The latter implies that $X$ is insensitive to all elements of $Y - X$. 

Conversely, assume that $X$ is insensitive to all $x \in Y - X$ and $Y$ is sensitive to some $x \in X - Y$. Then, $f(X \cup Y) = X$ by Lemma~\ref{key lemma} and hence $(X,Y)$ is a witness pair.  \hfill $\Box$

\vskip0.2cm

We are now ready for the algorithm. 
The algorithm has two phases. In a preprocessing phase, we determine whether $\Y$ is coherent, complete,  and, most importantly, compute the Boolean flags $\sens(x,X)$ which is true if $X \in \Y$ is sensitive to $x$. 

In the main computation, we search for the first witness to non-substitutability. We iterate over the elements of $X$ of $\Y$ in increasing order. Assume that there is no witness 
$(Z,\cdot )$ with $Z \succ X$. We then iterate over the $Y \in \Y$ with $X \succ Y$ and use Theorem~\ref{main theorem} to determine whether $(X,Y)$ is a witness pair.

The most expensive task of the 
first phase is the construction of the Boolean matrix $\sens$ of size $|U|\times N$.  
Since  an inclusion test needs $O(|U|)$ time, the overall time 
is therefore $O(|U|^2\cdot N^2)$.
 The time complexity of the second phase is  $O(|U|\cdot N^2)$ (the $|U|$ factor is given by the inspection of the Boolean matrix $\sens$ in order to apply Theorem~\ref{main theorem}).

By Theorems \ref{prop:thm1} and \ref{main theorem} and the above discussion, the following corollary  holds.

\begin{corollary}\label{prop:thm2} The algorithm in Figure \ref{fig:as1} tests in $O(|U|^2\cdot N^2)$ time 
if a given preference list of size $N$ over an universe $U$  is  substitutable.
\end{corollary}

 \noindent \textbf{Remarks.} 
The $O(N)$ speed-up over the existing algorithms is significant since (as we noted after the definition of complete lists) $N$ is exponential in the size of the largest member of $\Y$. The algorithm in \cite{azizbrilharr13} also applies to weak preferences. We leave it as an open problem whether this also holds for our algorithm.
 

\bibliography{omega}
\end{document}